\documentclass{article}

\PassOptionsToPackage{numbers, compress}{natbib}
\usepackage[final]{nips_2017}

\usepackage[utf8]{inputenc} 
\usepackage[T1]{fontenc}    
\usepackage{hyperref}       
\usepackage{url}            
\usepackage{booktabs}       
\usepackage{amsfonts}       
\usepackage{nicefrac}       
\usepackage{microtype}      
\usepackage{graphics}
\usepackage{epsfig,epstopdf,subfigure}

\title{Deep Factorization for Speech Signal}

\author{
  Dong Wang, Lantian Li, Ying Shi, Yixiang Chen, Zhiyuan Tang \\
  Center for Speech and Language Technologies, Tsinghua University \\
  \texttt{\{wangdong99,lilt13\}@mails.tsinghua.edu.cn} \\
}

\begin{document}

\maketitle

\begin{abstract}

  Speech signals are complex intermingling of various informative factors, and this information blending
  makes decoding any of the individual factors extremely difficult.
  A natural idea is to factorize each speech frame into independent factors, though it turns out to be
  even more difficult than decoding each individual factor. A major encumbrance
  is that the speaker trait, a major factor in speech signals,
  has been suspected to be a long-term distributional pattern and so not identifiable at the frame level.
  In this paper, we demonstrated that the speaker factor is also a short-time spectral pattern and
  can be largely identified with just a few frames using a simple deep neural network (DNN).
  This discovery motivated a \emph{cascade deep factorization} (CDF) framework that infers speech factors in a sequential
  way, and factors previously inferred are used as conditional variables when inferring other factors. Our experiment
  on an automatic emotion recognition (AER) task demonstrated that this approach can effectively factorize speech signals,
  and using these factors, the original speech spectrum can be recovered with high accuracy. This factorization and
  reconstruction approach provides a novel tool for many speech processing tasks.

\end{abstract}

\section{Introduction}

Speech signals are mysterious and fascinating: within just one dimensional vibration, very rich information is represented,
including linguistic content, speaker trait, emotion, channel and noise.
Scientists have worked for several decades to
decode speech, with different goals that focus on different informative factors within the signal. This leads to
a multitude of speech information processing tasks, where automatic speech recognition (ASR) and speaker recognition (SRE) are among the
most important~\cite{benesty2007springer}. After decades of research, some tasks have been addressed pretty well, at least with
large amounts of data, e.g., ASR and SRE, while others remain difficult, e.g., automatic emotion recognition (AER)~\cite{el2011survey}.

A major difficulty of speech processing resides in the fact that multiple informative factors are intermingled together,
and therefore whenever we decode for a particular factor, all other factors contribute as uncertainties. A natural idea
to deal with the information blending is to factorize the signal into independent informative factors, so that each task
can take its relevant factors.
Unfortunately, this factorization turns out to be very difficult, in fact more difficult than decoding for individual factors.
The main reason is that how the factors are intermingled to compose the speech signal and how they impact each other is far
from clear to the speech community, which makes designing a simple yet effective factorization formula nearly impossible.

As an example, the two most significant factors, linguistic contents and speaker traits, corresponding
to what has been spoken and who has spoken, hold a rather complex correlation. Here `significant factors'
refer to those factors that cause significant variations within speech signals.
Researchers have put much effort to factorize speech signals based on these two factors,
especially in SRE research.  In fact, most of the famous SRE techniques are based on factorization models,
including the Gaussian mixture
model-universal background model (GMM-UBM)~\cite{Reynolds00}, the joint factor analysis (JFA)~\cite{Kenny07}
and the i-vector model~\cite{dehak2011front}. With these models, the variation caused by the linguistic factor is
explained away, which makes the speaker factor easier to identity (infer).
Although significant success has been achieved, all these models assume
a linear Gaussian relation between the linguistic, speaker and other factors, which is certainly over simplified.
Essentially, they all perform shallow and linear factorization,
and the speaker factors inferred are \emph{long-term distributional patterns} rather than \emph{short-time spectral patterns}.

It would be very disappointing if the speaker factor is really a distributional pattern in nature, as it would
mean that speaker traits are too volatile to be identified from a short-time speech segment. If this is true, then it would be
hopeless to factorize speech signals into independent factors at the frame level, and for most speech processing tasks, we have to
resort to complex probabilistic models to collect statistics from long speech segments. This notion has in fact been subconsciously embedded into the thought process of many speech researchers, partly due to the brilliant success of probabilistic models on SRE.

Fortunately, our discovery reported in this paper demonstrated that the speaker trait is essentially a short-time spectral
pattern, the same as the linguistic content. We designed a deep neural network (DNN) that can learn speaker
traits pretty well from raw speech features, and demonstrated that with only a few frames, a very strong speaker factor
can be inferred.
Considering that the linguistic factor can be inferred from
a short segment as well~\cite{hinton2012deep}, our finding indicates that most of the
significant variations of speech signals can be well explained. Based on the explanation,
less significant factors are easier to be inferred. This has motivated a \emph{cascaded deep factorization}
(CDF) approach that factorizes speech signals in a sequential way: factors that are most significant
are inferred firstly, and other less significant factors are inferred subsequently, conditioned on the
factors that have been inferred. By this approach, speech signals can be factorized into
independent informative factors, where all the inferences are based on deep neural models.

In this paper, we apply the CDF approach to factorize emotional speech signals to linguistic contents,
speaker traits and emotion status. Our experiments on an AER task demonstrated that the CDF-based factorization
is highly effective. Furthermore, we show that the original speech signal can be reconstructed from these
three factors pretty well. This factorization and reconstruction has far-reaching implications and
will provide a powerful tool for many speech processing tasks.

\section{Speaker factor learning}
\label{sec:speaker}

In this section, we present a DNN structure that can learn speaker traits
at the frame level, as shown in Figure~\ref{fig:ctdnn}. This structure consists
of a convolutional (CN) component and a time-delay (TD) component,
connected by a bottleneck layer of $512$ units.
The convolutional component comprises two CN layers, each followed by a max-pooling layer.
The TD component comprises two TD layers, each followed by a P-norm layer.
The settings for the two components are shown in Figure~\ref{fig:ctdnn}.
A simple calculation shows that with this configuration, the length of the
effective context window is $20$ frames.
The output of the P-norm layer is projected into a feature layer that consists of $40$ units. The
activations of these units, after length normalization, form a speaker factor that
represents the speaker trait involved in the input speech segment. For
model training, the feature layer is fully connected to the output layer whose units correspond to
the speakers in the training data. The training is performed to optimize the cross-entropy objective
that aims to discriminate the training speakers based on the input frames. In our experiment,
the natural stochastic gradient descent (NSGD)~\cite{povey2014parallel} algorithm was employed for optimization. Once the DNN model has been trained, the $40$-dimensional
frame-level speaker factor can be read from the feature layer.
The speaker factors inferred by the DNN structure, as will be shown in the experiment, are highly
speaker-discriminative. This demonstrates that speaker traits are short-time
spectral patterns and can be identified at the frame level.

\begin{figure*}[htb]
    \centering
    \includegraphics[width=1\linewidth]{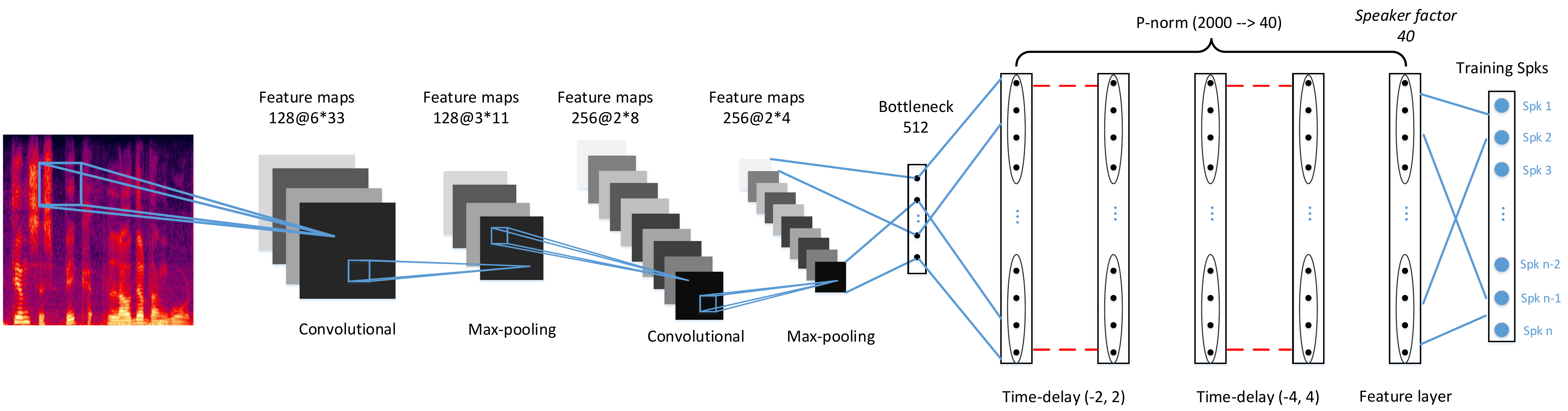}
    \caption{The DNN structure used for deep speaker factor inference.}
    \label{fig:ctdnn}
\end{figure*}

\section{Cascaded deep factorization}

Due to the highly complex intermingling of multiple informative factors, it is nearly impossible to
factorize speech signals by conventional linear factorization methods, e.g., JFA~\cite{Kenny07}.
Fortunately, the ASR research has demonstrated that the linguistic factor can be individually
inferred by a DNN structure, without knowing other factors. The previous section
further provides deep model that can infer the speaker factor.
We denote this single factor inference based on deep neural models by \emph{individual deep factorization} (IDF).

The rationality of the linguistic and speaker IDF is two-fold: firstly the linguistic and speaker factors are
sufficiently significant in speech signals, and secondly a large amount of training data is available.
It is the large-scale supervised learning that picks up the most task-relevant factors from
raw speech features, via the DNN architecture. For factors that are less significant or without sufficient
training data, IDF is simply not applicable. Fortunately, the successful inference of the linguistic
and/or the speaker factors may significantly simplify the inference of other speech factors,
as the largest variations within the speech signal have been explained away.
This has motivated a cascaded deep factorization (CDF) approach: firstly we infer a particular factor
by IDF, and then use this factor as a conditional variable to infer the second factor, and so on.
Finally, the speech signal will be factorized into a set of independent factors, each corresponding to a
particular task. The order of the inference can be arbitrary, but a good practice is that
factors that are more significant and with more training data should be inferred earlier, so that the
variation caused by these factors can be reliably eliminated when inferring the subsequent factors.

\begin{figure}[htb]
    \centering
    \includegraphics[width=0.5\linewidth]{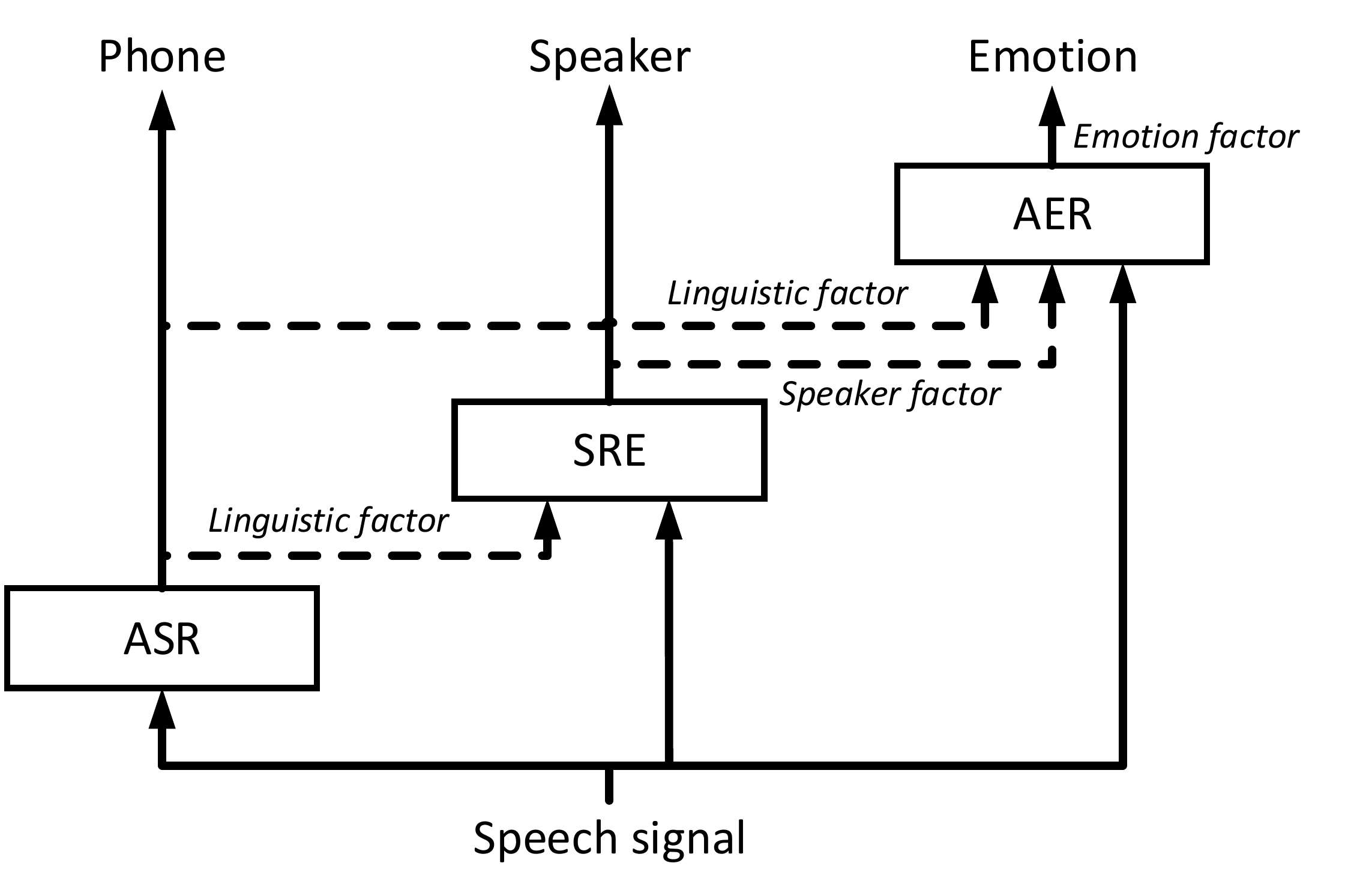}
    \caption{The cascaded deep factorization approach applied to factorize emotional speech into
    three factors: linguistic, speaker and emotion. }
    \label{fig:cascade}
\end{figure}

In this study, we apply the CDF approach to factorize emotional speech signals into three factors:
linguistic, speaker and emotion. Figure~\ref{fig:cascade} illustrates the architecture. Firstly
an ASR system is trained using word-labelled speech data.
The frame-level linguistic factor, which is in the form of phone posteriors in our study, is produced
from the ASR DNN, and is concatenated with the raw feature to train an SRE system.
This SRE system is used to produce the frame-level speaker factor, as discussed in the previous section.
The linguistic factor and the speaker factor are finally concatenated with the raw
feature to train an AER system, by which the emotion factor is read from the last hidden layer.

The CDF approach is fundamentally different from the conventional joint factorization approach,
e.g., JFA~\cite{Kenny07}. Firstly, CDF heavily relies on discriminative learning to discover task-related
factors, while conventional approaches are mostly generative models and the factors inferred are less
task-related. Secondly, CDF infers factors sequentially and can use different data resources for
different factors, while conventional approaches infer factors jointly
using a single multi-labelled database. Thirdly, CDF being a deep approach, can leverage various advantages
associated with deep learning (e.g., invariant feature learning), while most conventional approaches are mostly
based on shallow models.


\section{Spectrum reconstruction}
\label{sec:recovery}

A key difference between CDF and the conventional factor analysis~\cite{christopher2006pattern} is that
in CDF each factor is inferred individually, without any explicit constraint defined among the factors (e.g.,
the linear Gaussian relation as in JFA).
This on one hand is essential for a flexible factorization, but on the other hand, shuns an important
question: How these factors are composed together to produce the speech signal?

\begin{figure}[htb]
    \centering
    \includegraphics[width=0.85\linewidth]{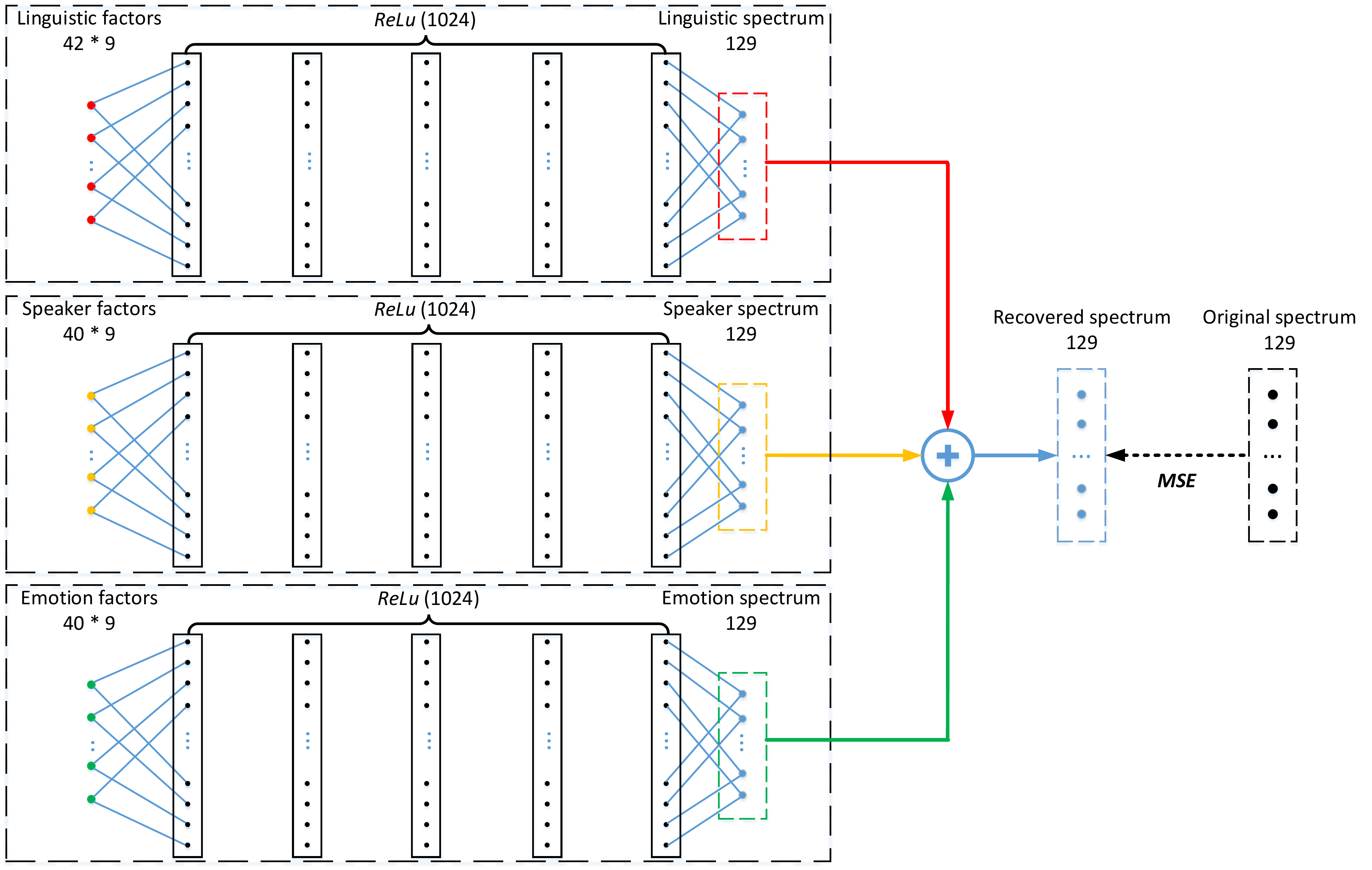}
    \caption{The architecture for spectrum reconstruction.}
    \label{fig:recovery}
\end{figure}

To answer this question, we reconstruct the spectrum using the CDF-inferred factors. Define
the linguistic factor $q$, the speaker factor $s$, and the emotion factor $e$.
For each speech frame, we try to use these three factors to
recover the spectrum $x$. Assuming they are convolved, the reconstruction is in the form:

\[
   ln (x) = ln \{f(q)\} + ln \{g(s)\} + ln \{h(e)\} + \epsilon
\]

\noindent where $f$, $g$, $h$ are the non-linear recovery function for $q$, $s$ and $e$ respectively,
each implemented as a DNN. $\epsilon$ represents the residual which is assumed to be Gaussian.
This reconstruction is illustrated in Figure~\ref{fig:recovery}, where all the spectra are in the log domain.

\section{Related work}

The idea of learning speaker factors was motivated by Ehsan et al~\cite{ehsan14}, who employed a vanilla DNN to
learn frame-level representations of speakers. These representations, however, were rather weak and
did not perform well on SRE tasks. Since then, various DNN structures were investigated,
e.g., RNN by Heigold~\cite{heigold2016end}, CNN by Zhang~\cite{zhang2017end} and NIN by
Snyder~\cite{snyderdeep16} and Li~\cite{li2017}. These diverse investigations demonstrated
reasonable performance, however most of them were based on
an end-to-end training, seeking for better performance on speaker verification, rather than
factor learning.

The CDF approach is also related to the phonetic DNN i-vector approach proposed by
Lei~\cite{lei2014novel} and Kenny~\cite{Kenny14}, where the linguistic factor (phonetic posteriors)
is firstly inferred using an ASR system, which is then used as an auxiliary knowledge to infer the speaker
factor (the i-vector). In CDF, the second stage linear Gaussian inference (i-vector inference)
is replaced by a more complex deep speaker factorization.

Finally, the CDF approach is related to multi-task learning~\cite{caruana1997multitask} and transfer
learning~\cite{pan2010survey,wang2015transfer}. For example, Senior et al.~\cite{senior2014improving}
found that involving the speaker factor in the input feature improved ASR system. Qin~\cite{qian2016neural}
and Li et al.~\cite{li2015modeling} found that ASR and SRE systems can be trained jointly, by borrowing
information from each other. This idea was recently studied more systematically by
Tang et al.~\cite{tang2017collaborative}. All these approaches focus on linguistic and speaker factors that are
mostly significant. The CDF, in contrast, treats these significant factors
as conditional variables and focuses more on less significant factors.

\section{Experiment}

In this section, we first present the data used in the experiments, then report the results of speaker factor
learning. The CDF-based emotional speech factorization and reconstruction will be also presented.

\subsection{Database}

\textbf{ASR database}: The \emph{WSJ} database was used to train the ASR system. The training set
is the official \emph{train\_si284} dataset, composed of $282$ speakers and $37,318$ utterances,
with about $50$-$155$ utterances per speaker. The test set contains three datasets (\emph{devl92, eval92 and eval93}),
including $27$ speakers and $1,049$ utterances in total.

\textbf{SRE database}: The \emph{Fisher} database was used to train the SRE systems. The training set consists of $2,500$ male and $2,500$ female speakers, with $95,167$ utterances randomly selected from the \emph{Fisher} database, and each speaker has about $120$ seconds of speech signals. It was used for training the UBM, T-matrix and LDA/PLDA models of an i-vector baseline system, and the DNN model proposed in Section~\ref{sec:speaker}. The test set consists of $500$ male and $500$ female speakers randomly selected from the \emph{Fisher} database. There is no overlap between the speakers of the training set and the evaluation set. For each speaker, $10$ utterances (about $30$ seconds in total) are used for enrollment and the rest for test. There are $72,989$ utterances for evaluation in total.

\textbf{AER database}: The \emph{CHEAVD} database~\cite{bao2014building} was used to train the AER systems.
This database was selected from Chinese movies and TV programs and used as the standard database for the multimodal emotion recognition challenge (MEC 2016)~\cite{li2016mec}.
There are $8$ emotions in total: Happy, Angry, Surprise, Disgust, Neutral, Worried, Anxious and Sad.
The training set contains $2,224$ utterances and the evaluation set contains $628$ utterances.

Note that WSJ and CHEAVD datasets are in 16kHz sampling rate, while the Fisher corpus is in 8kHz format.
All the 16kHz speech signals were down-sampled to 8kHz to ensure data consistency.

\subsection{ASR baseline}

We first build a DNN-based ASR system using the WSJ database. This system will be used to produce the linguistic factor
in the following CDF experiments. The Kaldi toolkit~\cite{povey2011kaldi} is used to train the DNN model,
following the Kaldi WSJ s5 nnet recipe. The DNN structure consists of $4$ hidden layers, each containing $1,024$ units.
The input feature is Fbanks, and the output layer discriminates $3,383$ GMM pdfs. With the official 3-gram language model,
the word error rate (WER) of this system is $9.16$\%. The linguistic factor is represented by
$42$-dimensional phone posteriors, derived from the output of the ASR DNN.

\subsection{Speaker factor learning}

In this section, we experiment with DNN structure proposed in Section~\ref{sec:speaker} to learn
speaker factors. Two models are investigated: one follows the architecture shown
in Figure~\ref{fig:ctdnn}, where only the raw features (Fbank) comprise the input; the other
model uses both the raw features \emph{and} the linguistic factors produced by the ASR system. Put it in another
way, the first model is trained by IDF, while the second model is trained by CDF. The Fisher database
is used to train the model. The 40-dimensional frame-level speaker factors are read out from the last hidden
layer of the DNN structure.

\textbf{Visualization}

\begin{figure}[htb]
    \centering
    \includegraphics[width=0.8\linewidth]{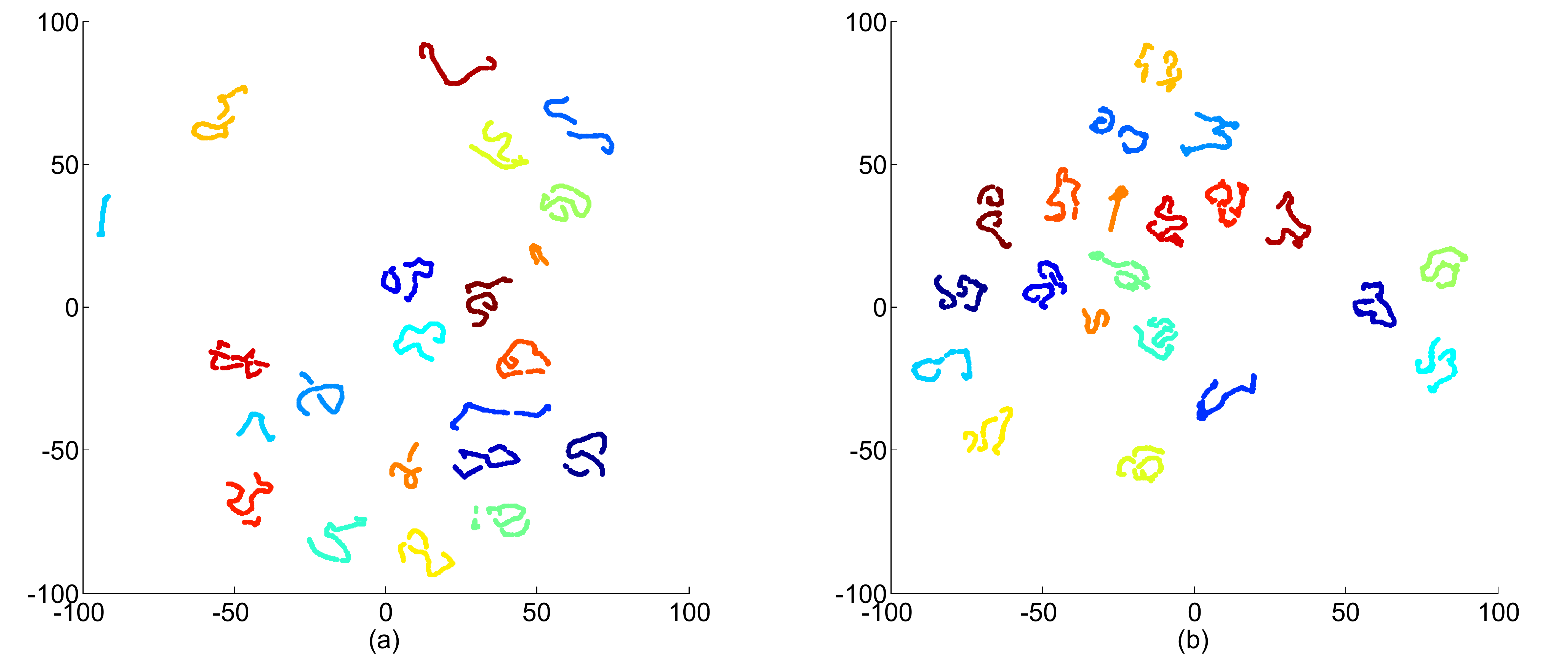}
    \caption{Frame-level speaker factors produced by (a) IDF DNN and (b) CDF DNN and plotted by t-SNE, with each color
    representing a speaker.}
    \label{fig:tsne}
    \end{figure}

The discriminative capability of the speaker factor can also be examined by projecting the
feature vectors to a 2-dimensional space using t-SNE~\cite{saaten2008}. We select $20$ speakers and draw the frame-level speaker factors of an utterance
for each speaker.
The results are presented in Figure~\ref{fig:tsne} where plot (a) draws the factors generated by the IDF model,
and (b) draws the factors generated by the CDF model.
It can be seen that the learned speaker factors are very discriminative, and involving the linguistic
factor by CDF indeed reduces the within-speaker variation.

\textbf{SRE performance}

The quality of the speaker factors can be also evaluated by various speaker recognition tasks, i.e., speaker
identification task or speaker verification task.
In both tasks, the utterance-level speaker vector is derived
by averaging the frame-level speaker factors. Following the convention of Ehsan et al~\cite{ehsan14},
the utterance-level representations derived from DNN are called \emph{d-vectors}, and accordingly the SRE system
is called a \emph{d-vector system}.

For comparison, an i-vector baseline is also constructed using the same database. The model is a
typical linear Gaussian factorization model and it has been demonstrated to produce state-of-the-art
performance in SRE~\cite{dehak2011front}. In our implementation, the UBM is composed of $2,048$
Gaussian components, and the dimensionality of the i-vector space is set to $400$. The system is trained following the Kaldi SRE08 recipe.

We report the results on the identification task, though similar observations were obtained
on the verification task.
In the identification task, a matched speaker is identified given a
test utterance. With the i-vector (d-vector) system, each enrolled speaker is represented by
the i-vector (d-vector) of their enrolled speech, and the i-vector (d-vector) of the test speech
is derived as well. The identification is then conducted by finding the speaker whose enrolled i-vector (d-vector)
is nearest to that of the test speech.
For the i-vector system, the popular PLDA model~\cite{Ioffe06} is used to measure the similarity between
i-vectors; for the d-vector system, the simple cosine distance is used.

The results in terms of the Top-1 identification rate (IDR) are shown in Table~\ref{tab:id-short}.
In this table, `C(30-20f)' means the test condition where the enrollment speech is $30$ seconds, while the test
speech is $20$ frames. Note that $20$ frames is just the length of the effective context window of the
speaker DNN, so only a single speaker factor is used in this condition. From these results,
it can be observed that the d-vector system performs much better than the i-vector baseline, particularly with
very short speech segments. Comparing the IDF and CDF results, it can be seen that the CDF approach that
involves phone knowledge as the conditional variable greatly improves the d-vector system in the short speech
segment condition.
Most strikingly, with only $20$ frames of speech (0.3 seconds), $47.63$\% speakers
can be correctly identified from the $1,000$ candidates by the simple nearest neighbour search.
This is a strong evidence that speaker traits are short-time spectral patterns and can be
effectively learned at the frame level.

    \begin{table}[htb!]
    \begin{center}
      \caption{The \textbf{\emph{Top-1}} IDR(\%) results on the short-time speaker identification with the i-vector and two d-vector systems.}
      \label{tab:id-short}
          \begin{tabular}{|c|l|c|c|c|c|}
            \hline
            \multicolumn{2}{|c|}{}                 &\multicolumn{3}{c|}{IDR\%}\\
            \hline
               Systems              &  Metric    &   S(30-20f) &    S(30-50f)   &   S(30-100f) \\
           \hline
              i-vector               &    PLDA     &   5.72    &    27.77      &    55.06      \\
               d-vector (IDF)             &    Cosine   &   37.18   &    51.24 &    \textbf{65.31}      \\
               d-vector (CDF)        &    Cosine   & \textbf{47.63}   & \textbf{57.72}      &    64.45      \\
           \hline
          \end{tabular}
      \end{center}
   \end{table}

\subsection{Emotion recognition by CDF}

In the previous experiment we have partially demonstrated the CDF approach with the speaker factor learning task.
This section provides further evidence with an emotion recognition task.
For that purpose, we first build a DNN-based AER baseline. The DNN model consists of $6$ time-delay
hidden layers, each containing $200$ units. After each TD layer, a P-norm layer reduces the dimensionality
from $200$ to $40$. The output layer comprises $8$ units, corresponding to the $8$ emotions
in the database. This DNN model produces frame-level emotion posteriors. The utterance-level posteriors
are obtained by averaging the frame-level posteriors, by which the utterance-level emotion decision is achieved.

Three CDF configurations are investigated, according to which factor is used as the conditional:
the linguistic factor (+ ling.), the speaker factor (+ spk.) and both (+ ling. \& spk.).
The results are evaluated in two metrics:
the identification accuracy (ACC) that is the ratio of the correct identification on all emotion categories;
the macro average precision (MAP) that is the average of the ACC on each of the emotion category.

The results on the training data are shown in Table~\ref{tab:cdf}, where the ACC and MAP values
on both the frame-level (fr.) and the utterance-level (utt.) are reported. It can be seen that with the conditional
factors involved, either the linguistic factor or the speaker factor, the ACC and MAP values are
improved very significantly. The speaker factor seems provide more significant contribution, which
can be attributed to the fact that the emotion style of different speakers could be largely different.
With both the two factors involved, the AER performance is improved even further.
This clearly demonstrates that with the conditional factors considered, the speech signal
can be explained much better.

\begin{table}[thb!]
  \caption{\label{tab:cdf}{Accuracy (ACC) and macro average precision (MAP) on the training set.}}
  \centerline{
    \begin{tabular}{|l|c|c|c|c|c|}
      \hline
      Dataset          & \multicolumn{4}{|c|}{Training set} \\
      \hline
                       & ACC\% (fr.) & MAP\% (fr.) & ACC\% (utt.) & MAP\% (utt.) \\
      \hline
      Baseline         & 74.19 & 61.67 & 92.27  & 83.08               \\
      +ling.           & 86.34 & 81.47 & 96.94  & 96.63               \\
      +spk.            & 92.56 & 90.55 & 97.75  & 97.16               \\
      +ling. \& spk.      & \textbf{94.59} & \textbf{92.98} & \textbf{98.02}  & \textbf{97.34}   \\
      \hline
   \end{tabular}
  }
\end{table}

The results on the test data are shown in Table~\ref{tab:cdf-test}. Again, we observe a clear advantage
with the CDF training. Note that involving the two factors does not improve the
utterance-level results. This should be attributed to the fact that the DNN models are trained
using frame-level data, so may be not fully consistent with the metric of the utterance-level test.
Nevertheless, the superiority of the multiple conditional factors can be seen clearly from the frame-level
metrics.

\begin{table}[thb!]
  \caption{\label{tab:cdf-test}{Accuracy (ACC) and macro average precision (MAP) on the evaluation set.}}
  \centerline{
    \begin{tabular}{|l|c|c|c|c|c|}
      \hline
      Dataset      & \multicolumn{4}{|c|}{Evaluation set}\\
      \hline
                   & ACC\% (fr.) & MAP\% (fr.) & ACC\% (utt.) & MAP\% (utt.)  \\
      \hline
      Baseline     & 23.39 & 21.08 & 28.98  & 24.95  \\
      +ling.       & 27.25 & 27.68 & \textbf{33.12}  & \textbf{33.28}  \\
      +spk.        & 27.18 & 28.99 & 32.01  & 32.62  \\
      +ling. \& spk.  & \textbf{27.32} & \textbf{29.42} & 32.17  & 32.29  \\
      \hline
   \end{tabular}
  }
\end{table}

\subsection{Spectrum reconstruction}

\begin{figure}[htb]
    \centering
    \includegraphics[width=1\linewidth]{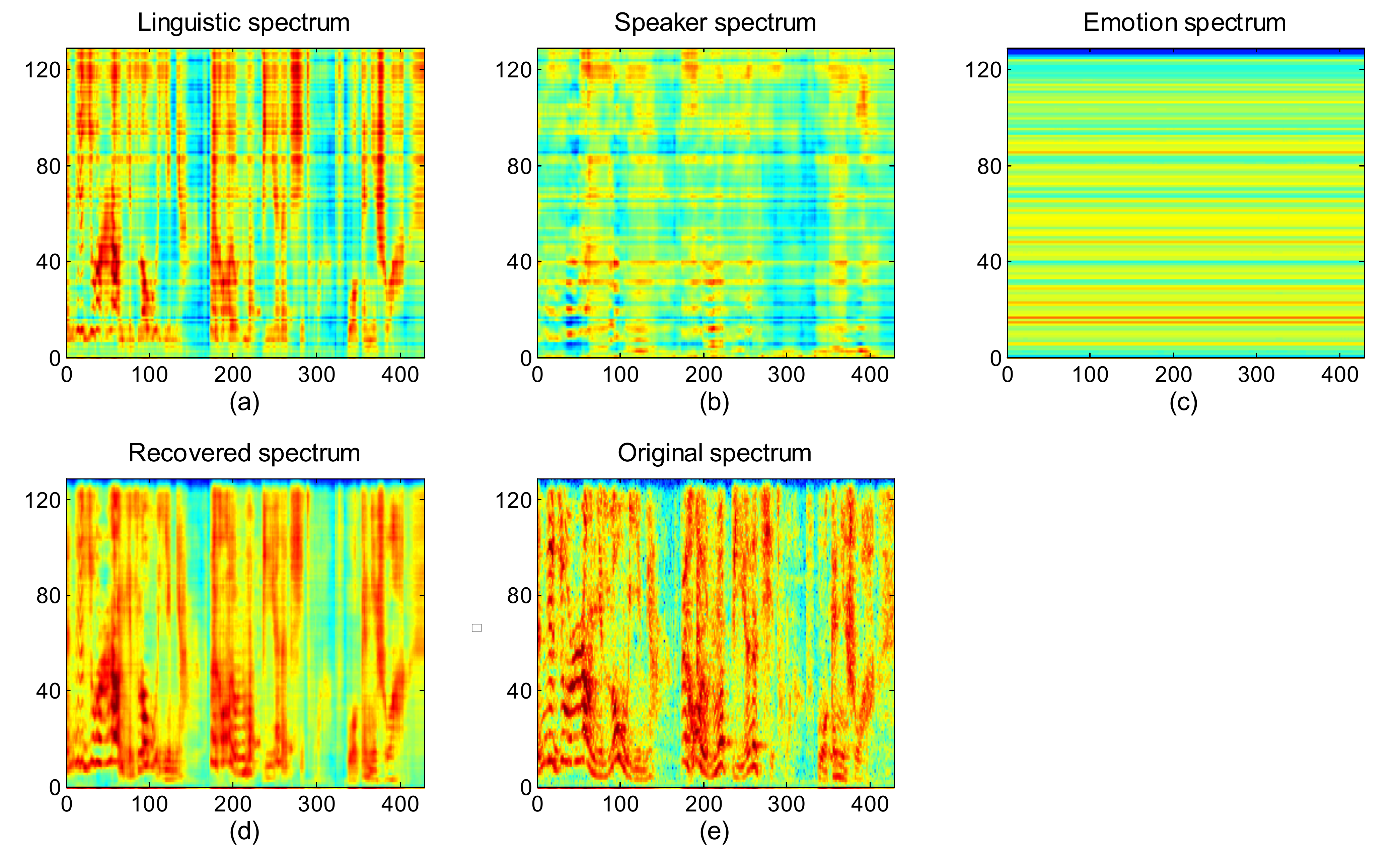}
    \caption{An example of spectrum reconstruction from the linguistic, speaker and emotion factors.}
    \label{fig:demo}
\end{figure}

In the last experiment, we use the linguistic factor, speaker factor and emotion factor to reconstruct
the original speech signal. The reconstruction model has been discussed in Section~\ref{sec:recovery}
and shown in Figure~\ref{fig:recovery}. This model is trained using the CHEAVD database. Figure~\ref{fig:demo}
shows the reconstruction of a test utterance in the CHEAVD database. It can be seen
that these three factors can reconstruct the spectrum patterns extremely well. This re-confirms that
the speech signal has been well factorized, and the convolutional reconstruction formula
is mostly correct. Finally, the three component spectra (linguistic, speaker, and emotion) are
highly interesting and all deserve extensive investigation. For example, the speaker spectrum may be
a new voiceprint analysis tool and could be very useful for forensic applications.

\section{Conclusions}

This paper has presented a DNN model to learn short-time speaker traits and a cascaded deep factorization (CDF)
approach to factorize speech signals into independent informative factors. Two interesting things
were found: firstly speaker traits are indeed short-time spectral patterns and can be
identified by deep learning from a very short speech segment; secondly speech signals can be well factorized
at the frame level by the CDF approach.
We also found that the speech spectrum can be largely reconstructed using deep neural models
from the factors that have been inferred by CDF, confirming the correctness of the factorization.
The successful factorization and reconstruction of speech signals has very important implications
and can find broad applications.
To mention several: it can be used to design very parsimonious speech codes, to change the speaker traits or
emotion in speech synthesis or voice conversion, to remove background noise, to embed audio watermarks.
All are highly interesting and are under investigation.

\subsubsection*{Acknowledgments}

Many thanks to Ravichander Vipperla from Nuance, UK for many valuable suggestions.

\newpage
\bibliographystyle{IEEEtran}
\bibliography{mybib}

\newpage
\section{Appendix A: Model details}

\subsection{ASR system}

The ASR system was built following the Kaldi WSJ s5 nnet recipe. The input
feature was 40-dimensional Fbanks, with a symmetric 5-frame window to splice neighboring frames.
It contained $4$ hidden layers, and each layer had $1,024$ units. The output layer consisted of
$3,383$ units, equal to the total number of pdfs of the GMM system trained following the WSJ s5 gmm recipe.
The language model was the WSJ official 3-gram model (`tgpr') that consists of $19,982$ words.

\subsection{SRE system}

The i-vector SRE baseline used 19-dimensional MFCCs plus the log energy as the primary feature. This primary
feature was augmented by its first and second order derivatives, resulting in a 60-dimensional feature vector.
The UBM was composed of $2,048$ Gaussian components, and the dimensionality of the i-vector space was $400$.
The entire system was trained using the Kaldi SRE08 recipe.

\begin{figure*}[htb]
    \centering
    \includegraphics[width=1\linewidth]{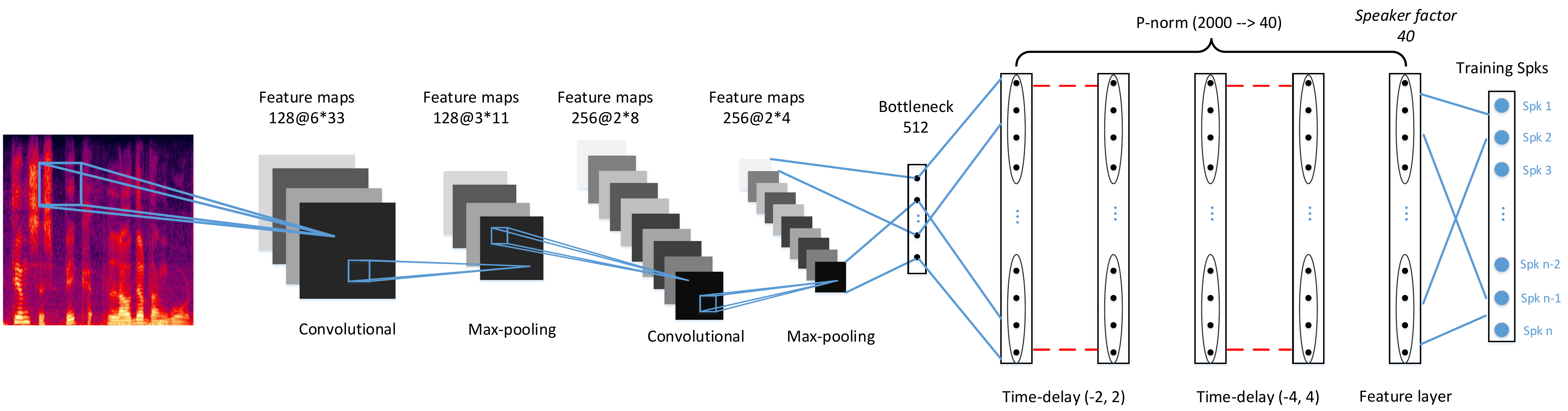}
    \caption{The DNN structure used for deep speaker factor inference.}
    \label{fig:ctdnn}
\end{figure*}

For the IDF d-vector system, the architecture was based on Figure~\ref{fig:ctdnn}. The input feature was
40-dimensional Fbanks, with a symmetric 4-frame window to splice the neighboring frames,
resulting in $9$ frames in total. The number of output units was $5,000$, corresponding to the number
of speakers in the training set.

\begin{figure*}[htb]
    \centering
    \includegraphics[width=1\linewidth]{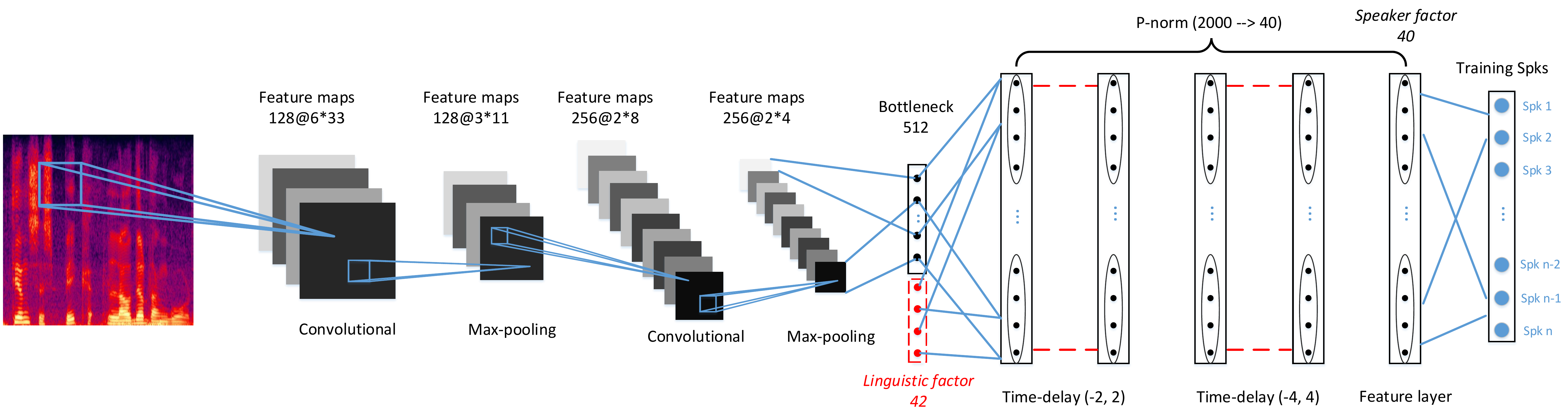}
    \caption{The CDF DNN structure used for deep speaker factor inference with the linguistic factor.}
    \label{fig:ctdnn-cdf}
\end{figure*}

For the CDF d-vector system, the linguistic factor in the form of $42$-dimensional phone posteriors was augmented to
the bottleneck layer, as shown in Figure~\ref{fig:ctdnn-cdf}.

\subsection{AER system}

The input feature of the DNN model of the AER baseline was 40-dimensional Fbanks, with a symmetric 4-frame
window to splice neighboring frames.
The time-delay component involving two time-delay layers was used to extend the temporal context,
and the length of the effective context window was $20$ frames.
It contained $6$ hidden layers, and each layer had $200$ units. With the P-norm activation, the dimensionality of
the output of the previous layer was reduced to $40$.

The definitions of ACC and MAP are given in Eqs.~\ref{eq:pi} - ~\ref{eq:acc}.

\begin{equation}
   P_i=\frac{TP_i}{TP_i + FP_i},
   \label{eq:pi}
\end{equation}

\begin{equation}
   MAP=\frac{1}{s} \times \sum\nolimits_{i=1}^{s}P_i,
   \label{eq:map}
\end{equation}

\begin{equation}
   ACC=\frac{\sum\nolimits_{i=1}^{s}TP_i}{\sum\nolimits_{i=1}^{s}{(TP_i + FP_i)}},
   \label{eq:acc}
\end{equation}

where $s$ denotes the number of emotion categories. $P_i$ is the precision of the $i^{th}$ emotion class.
$TP_i$ and $FP_i$ denote the number of correct classification and the number of error classification in the $i^{th}$ emotion class, respectively.

\subsection{Spectrum reconstruction}

The spectrum reconstruction is based on the following convolutional assumption:

\[
   ln (x) = ln \{f(q)\} + ln \{g(s)\} + ln \{h(e)\} + \epsilon
\]

\noindent where $f$, $g$, $h$ are the non-linear recovery function for $q$, $s$ and $e$ respectively,
each implemented as a DNN. $\epsilon$ represents the residual which is assumed to be Gaussian.

The DNN structure for the spectrum reconstruction consists of two parts:
A factor spectrum generation component and a spectrum convolution component.
The former generates component spectrum for each factor (e.g., $f(q)$, $g(s)$, $h(e)$),
and the latter composes the three component spectra together.

The dimensionalities of the linguistic, speaker and emotion factors are $42$, $40$ and $40$,
respectively. With a symmetric 4-frame window, the input dimensionalities of three spectrum-generation
components are $387$, $360$ and $360$, respectively.
Each spectrum-generation component involves $5$ hidden layers, each consisting of $1,024$ units and followed by
the ReLu (Rectified Linear Unit) non-linear activation function. The outputs of these three
spectrum-generation components are fed into the spectrum convolutional component, where the
reconstruction of the original spectrum is produced.

The MSE (Mean Squared Error) between the recovered spectrum and the original spectrum
is used as the training criterion. Note that the target spectrum is in the log domain, so the
convolution component is a simple addition.

\newpage
\section{Appendix B: Samples of spectrum reconstruction}

Here we give more examples to demonstrate the spectrum reconstruction.

\subsection{Training set}

\begin{figure}[!htb]
    \centering
    \includegraphics[width=1\linewidth]{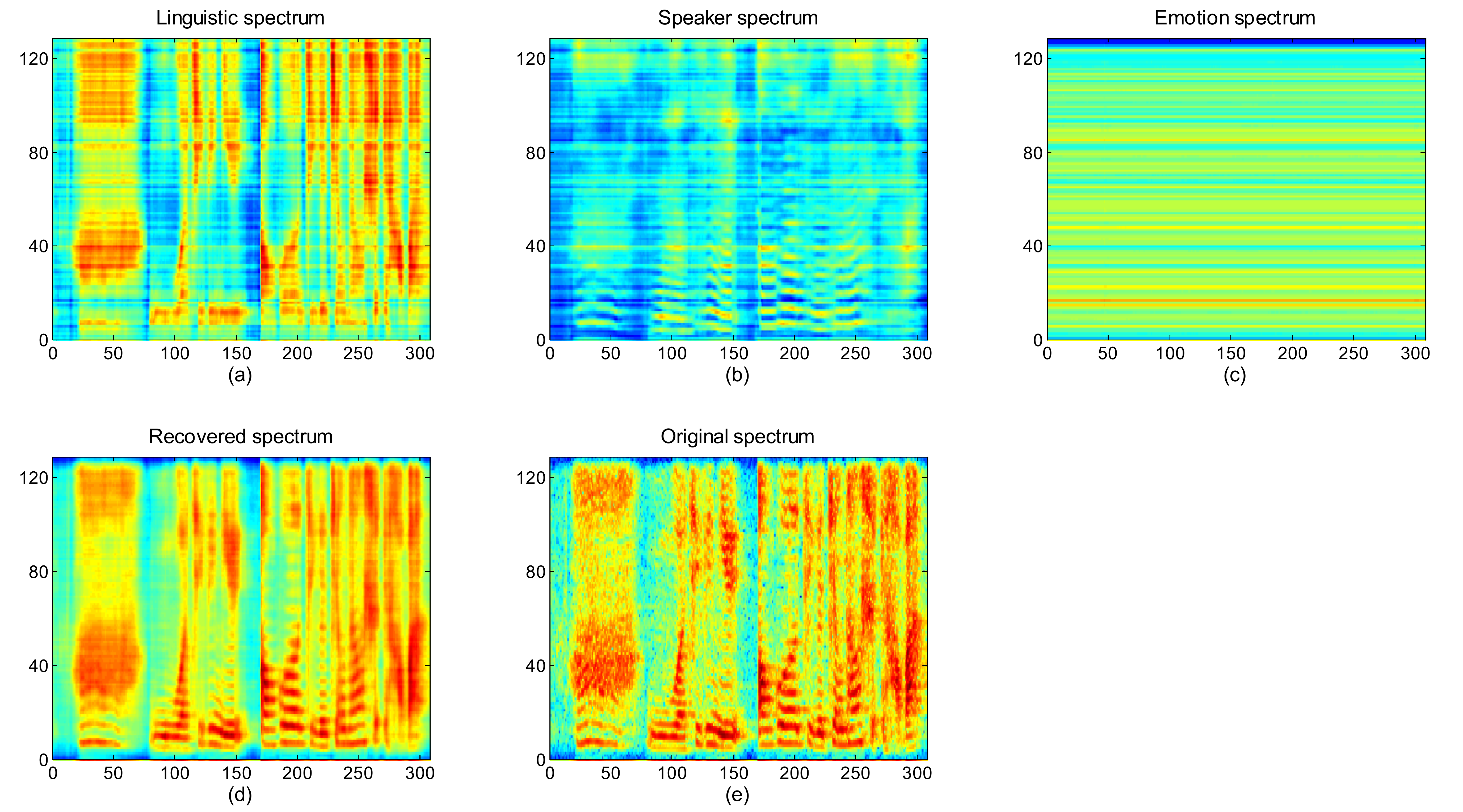}
    \caption{Training set (1).}
    \label{fig:demo-1}
\end{figure}

\begin{figure}[!htb]
    \centering
    \includegraphics[width=1\linewidth]{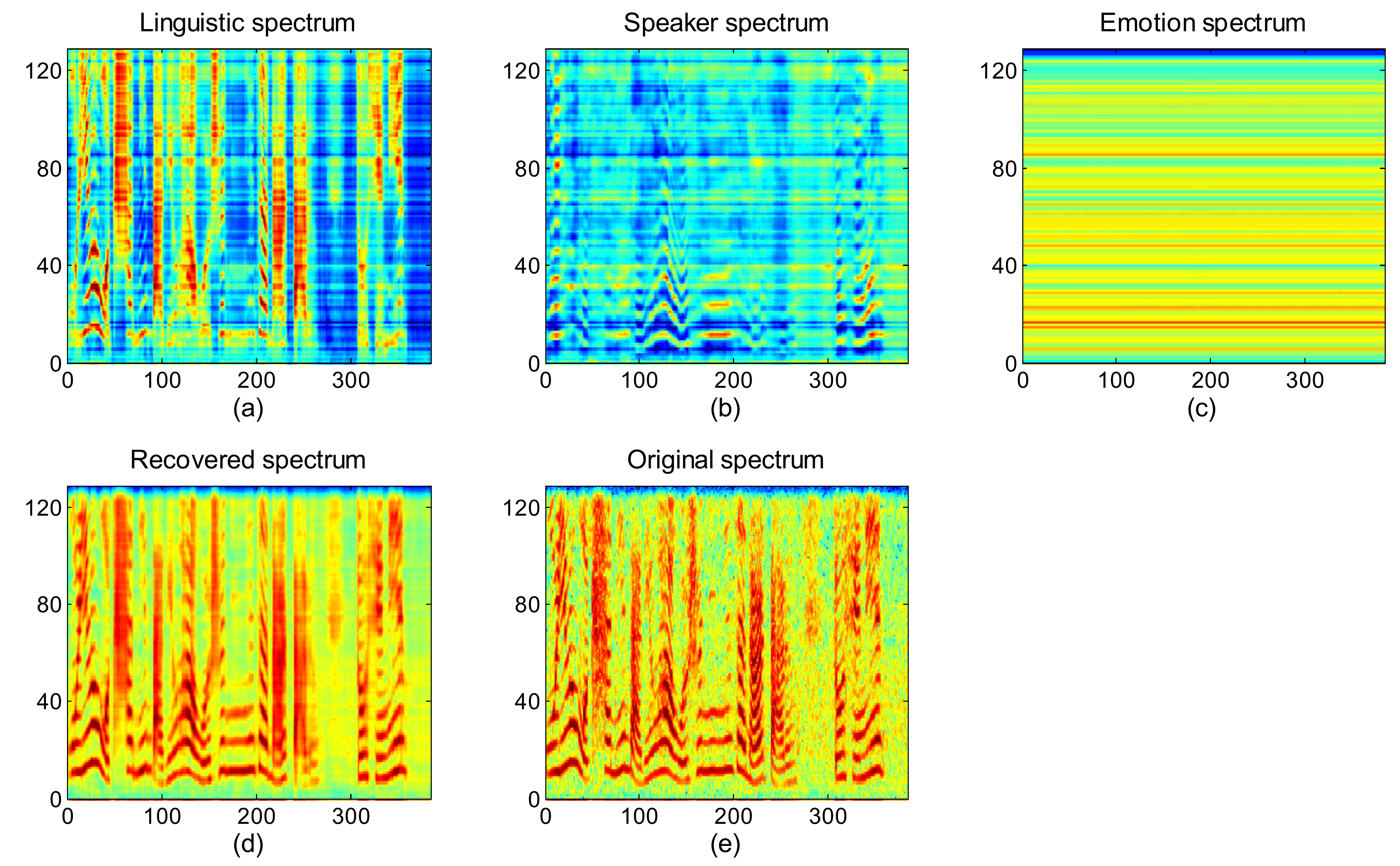}
    \caption{Training set (2).}
    \label{fig:demo-2}
\end{figure}

\newpage

\begin{figure}[!htb]
    \centering
    \includegraphics[width=1\linewidth]{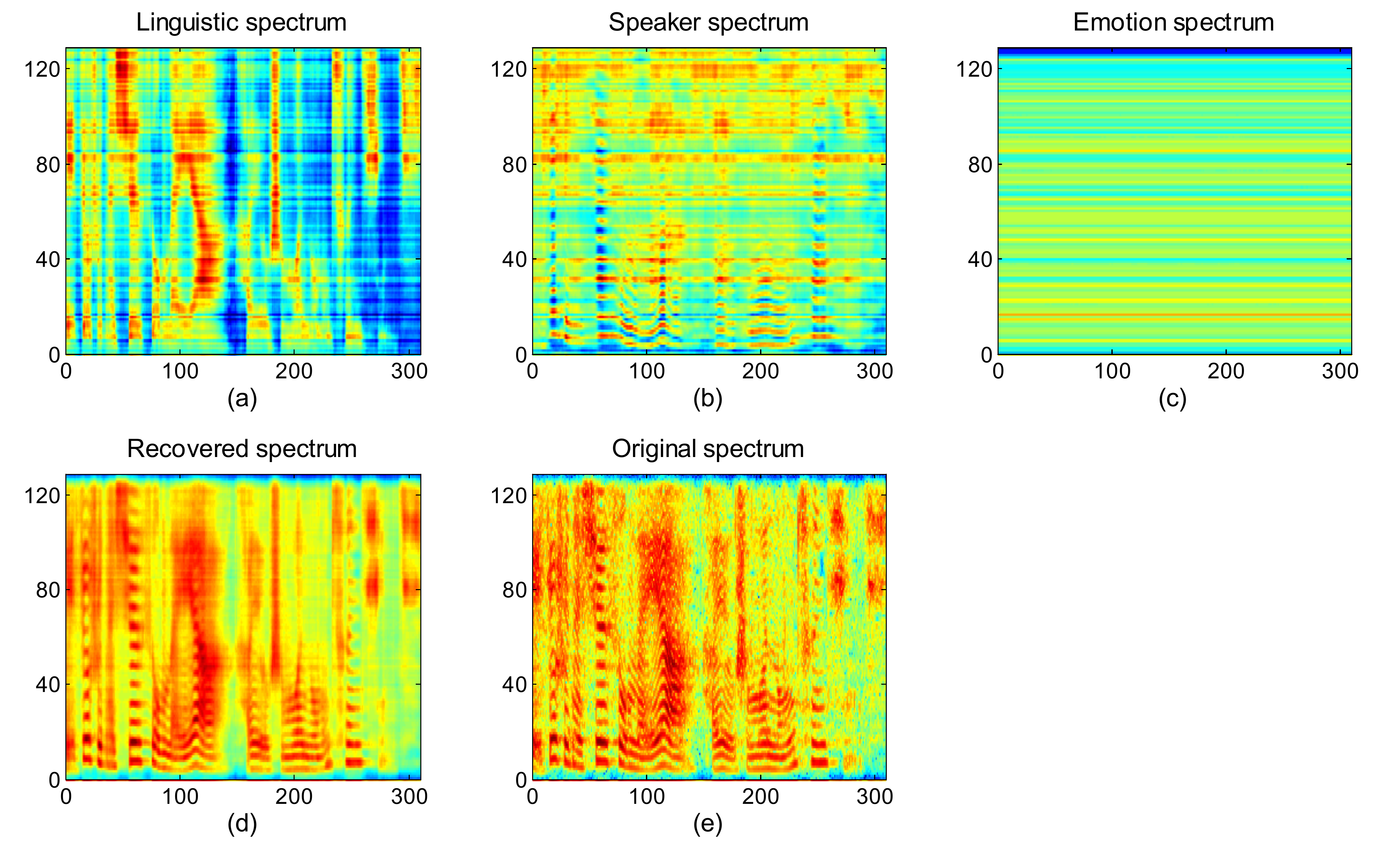}
    \caption{Training set (3).}
    \label{fig:demo-3}
\end{figure}

\begin{figure}[!htb]
    \centering
    \includegraphics[width=1\linewidth]{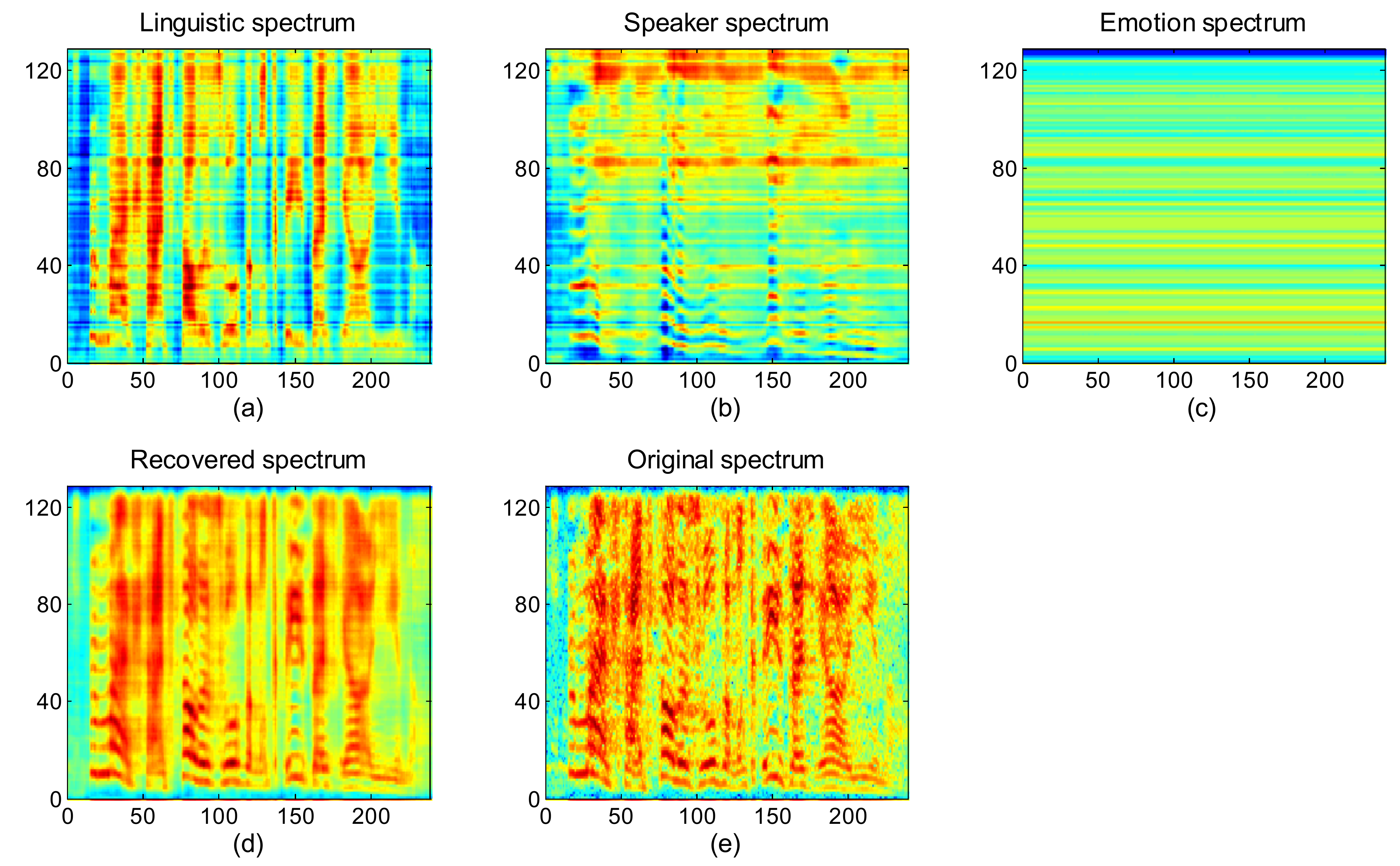}
    \caption{Training set (4).}
    \label{fig:demo-4}
\end{figure}

\newpage
\begin{figure}[!htb]
    \centering
    \includegraphics[width=1\linewidth]{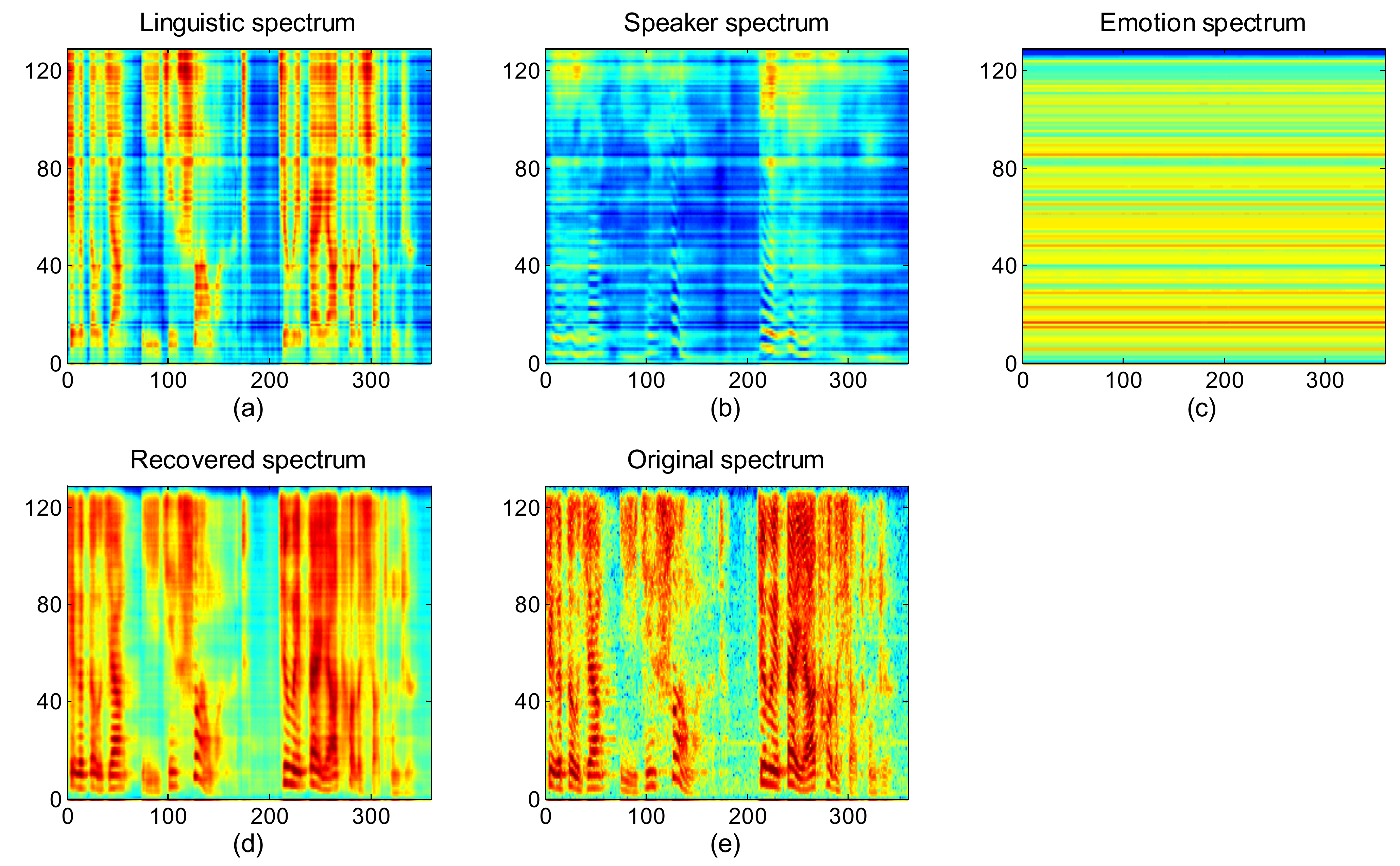}
    \caption{Training set (5).}
    \label{fig:demo-5}
\end{figure}

\subsection{Evaluation set}

\begin{figure}[!htb]
    \centering
    \includegraphics[width=1\linewidth]{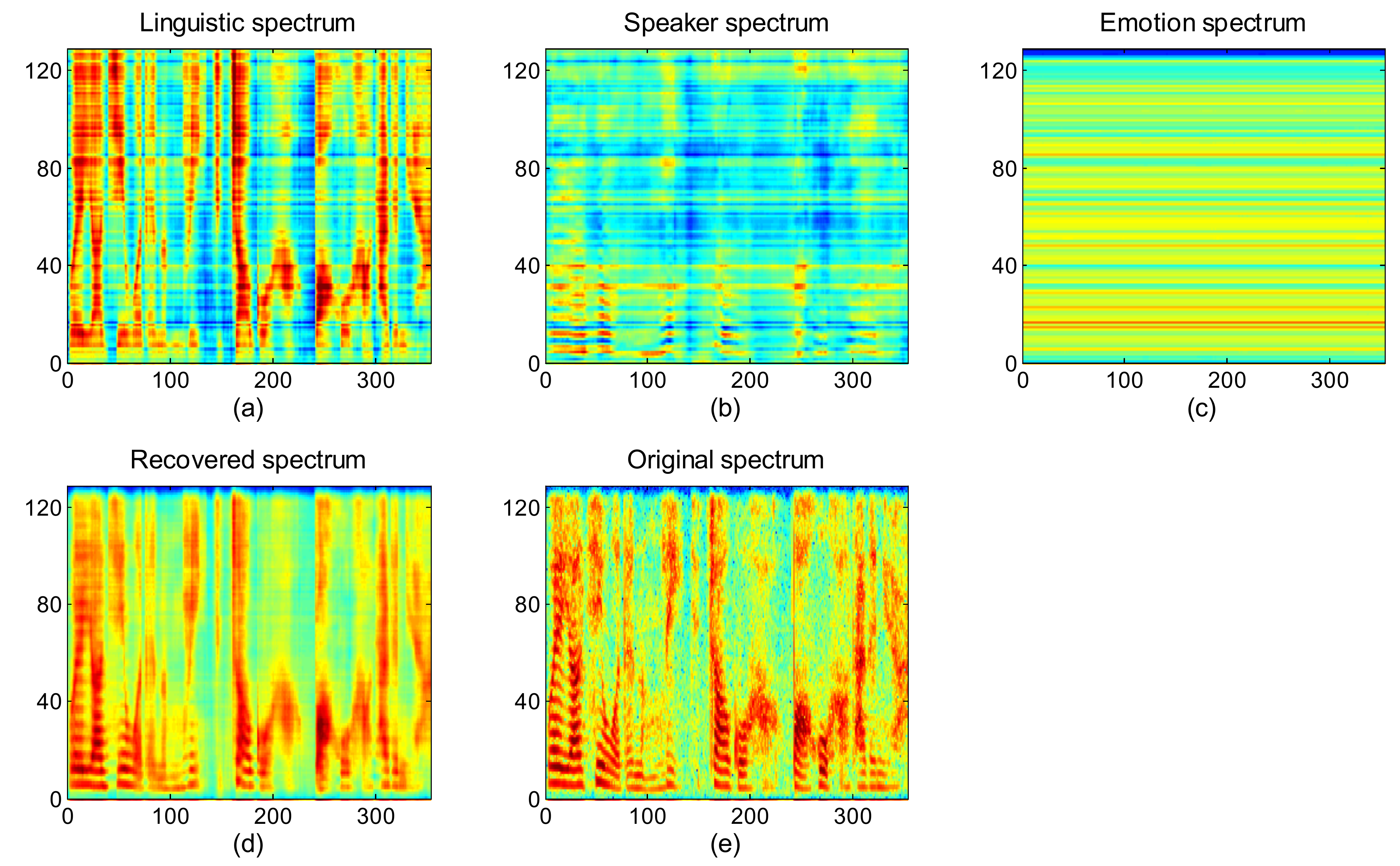}
    \caption{Evaluation set (1).}
    \label{fig:demo-6}
\end{figure}

\newpage

\begin{figure}[!htb]
    \centering
    \includegraphics[width=1\linewidth]{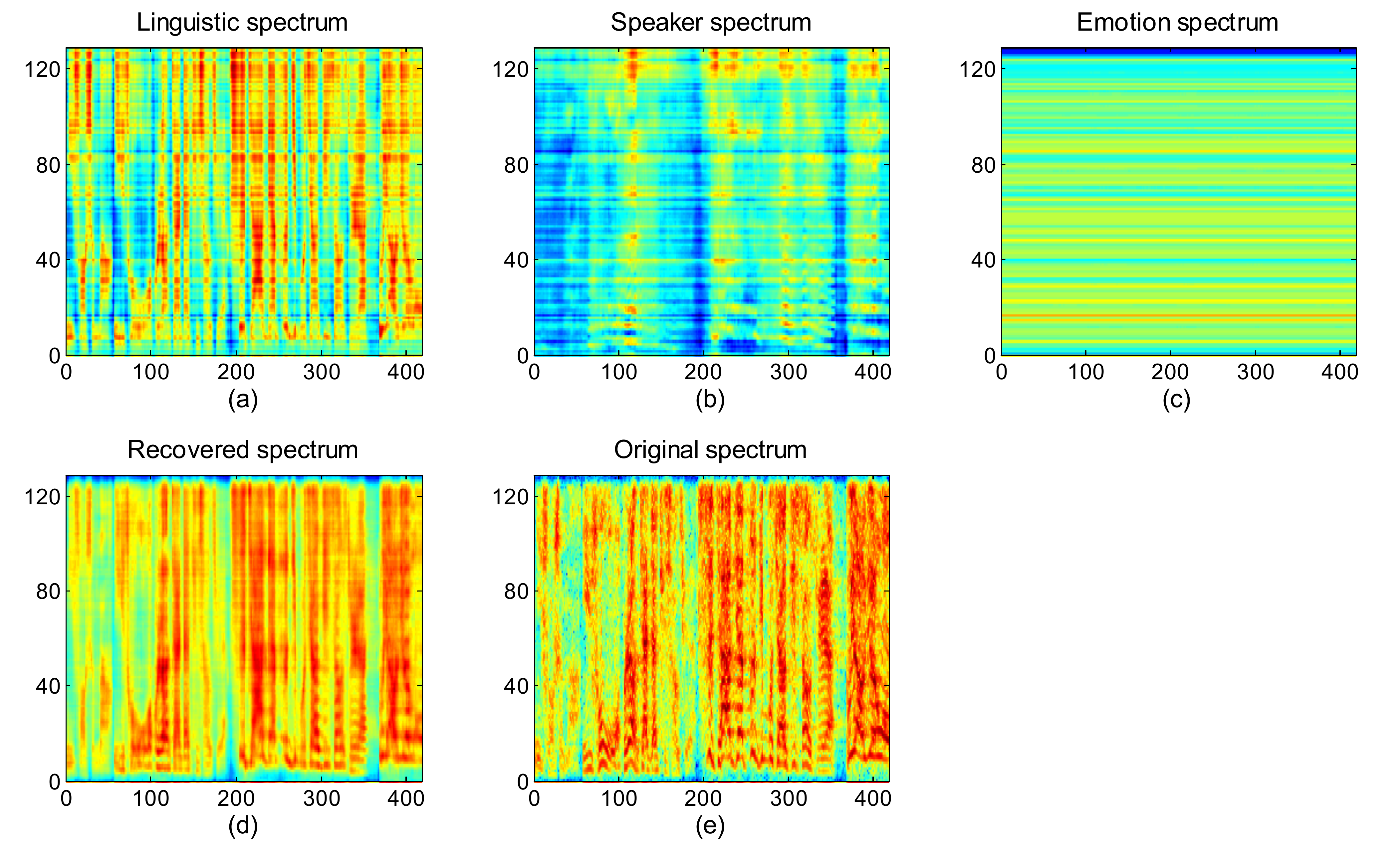}
    \caption{Evaluation set (2).}
    \label{fig:demo-7}
\end{figure}

\begin{figure}[htb]
    \centering
    \includegraphics[width=1\linewidth]{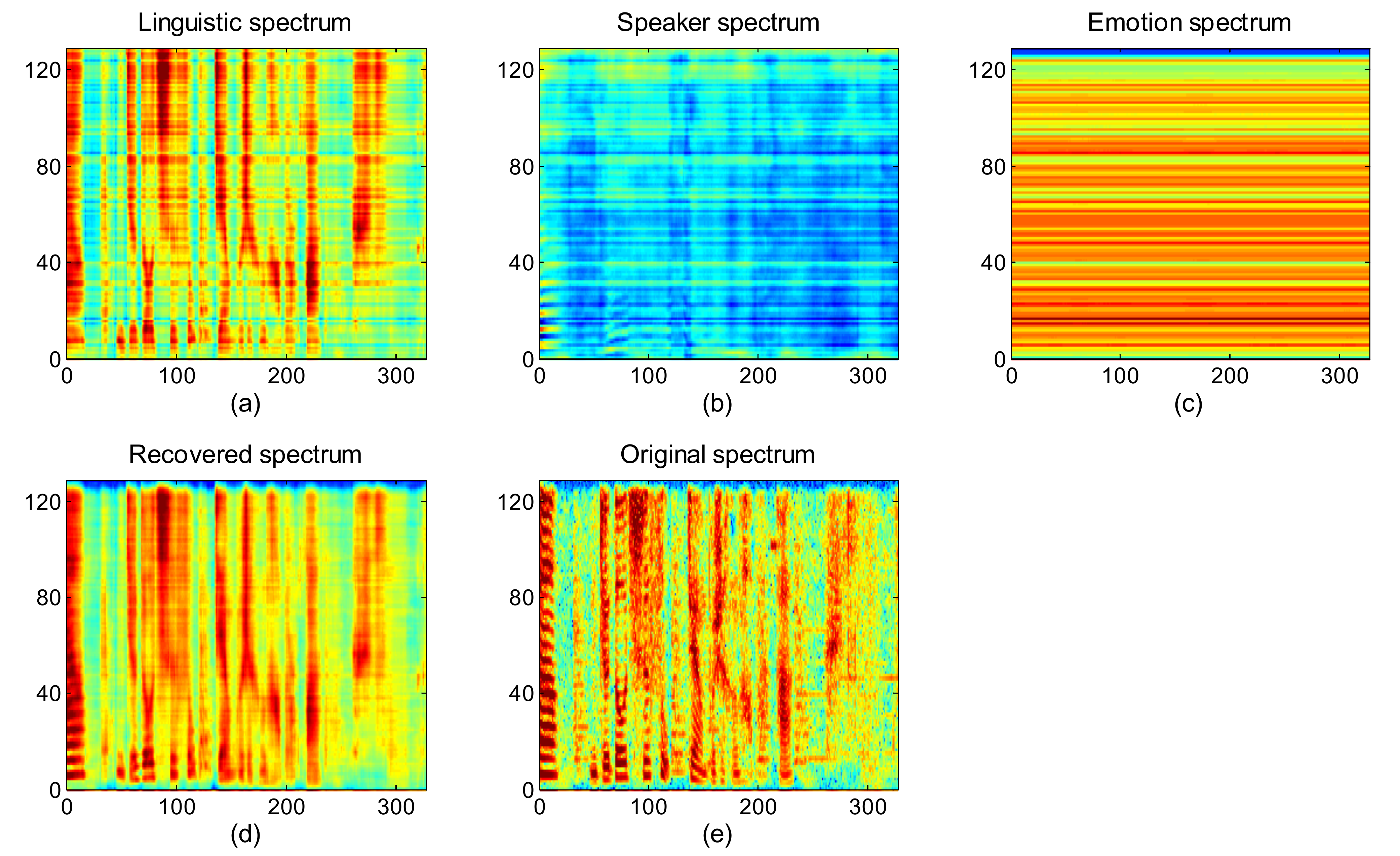}
    \caption{Evaluation set (3).}
    \label{fig:demo-8}
\end{figure}

\newpage
\begin{figure}[htb]
    \centering
    \includegraphics[width=1\linewidth]{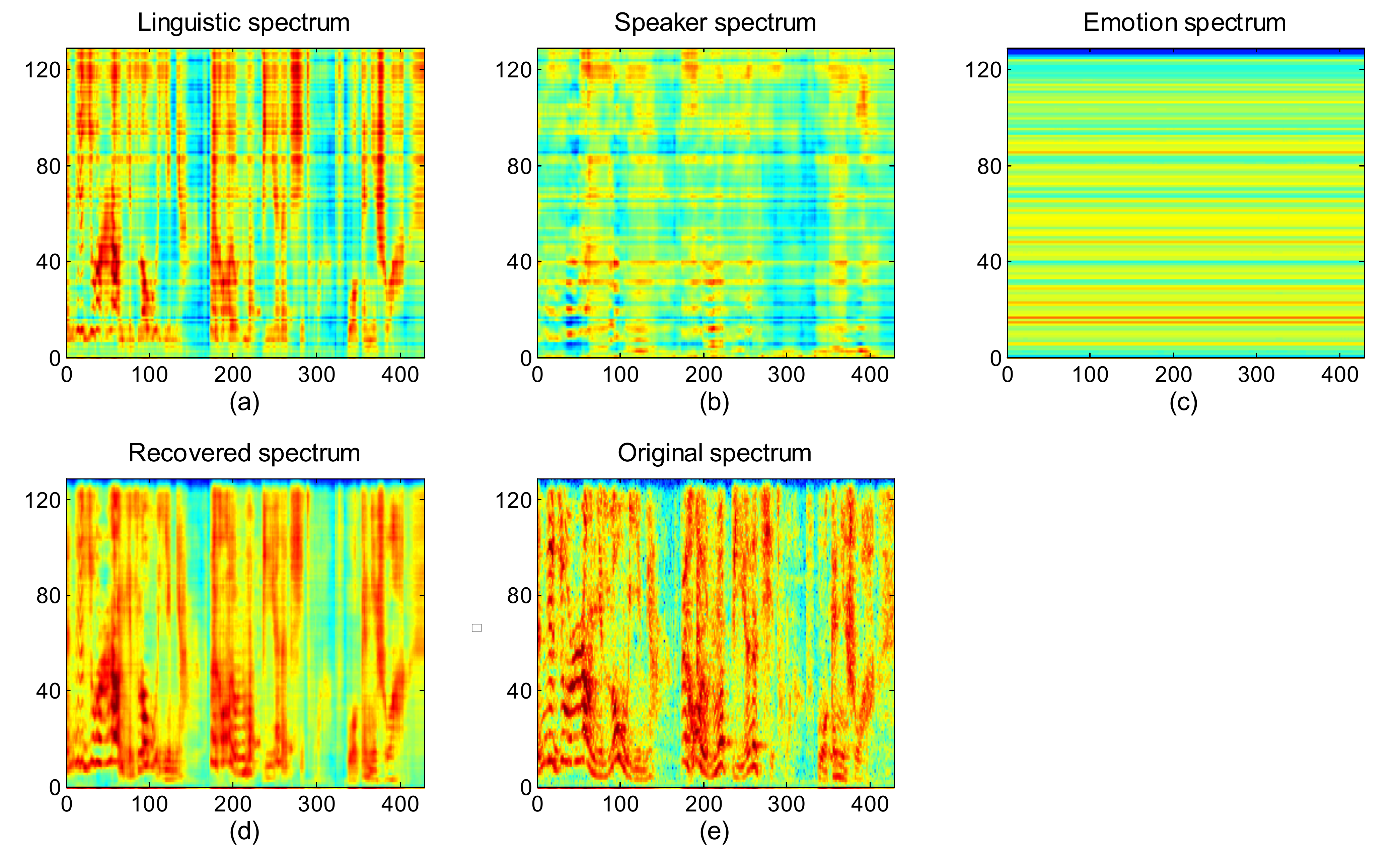}
    \caption{Evaluation set (4).}
    \label{fig:demo-9}
\end{figure}

\begin{figure}[htb]
    \centering
    \includegraphics[width=1\linewidth]{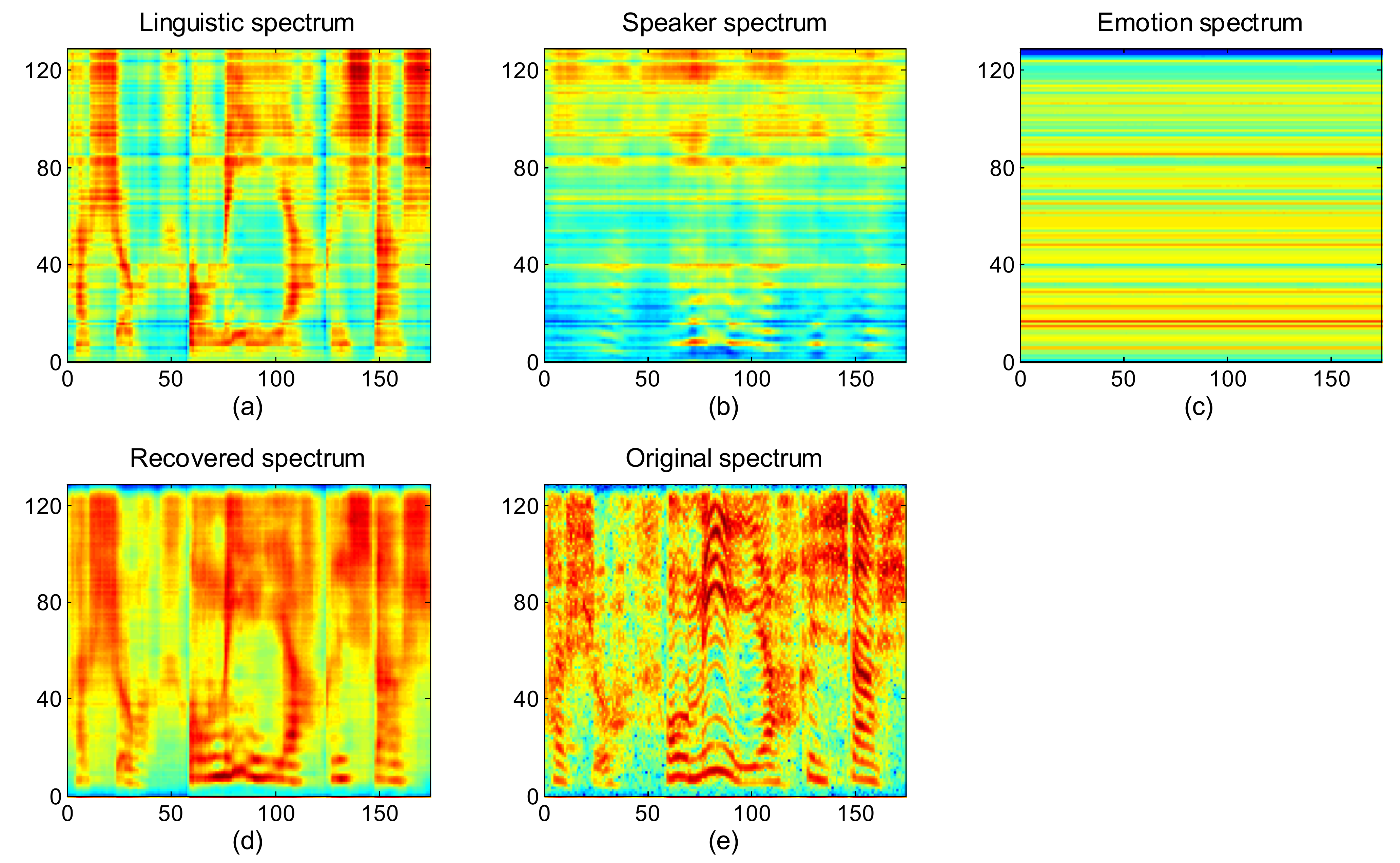}
    \caption{Evaluation set (5).}
    \label{fig:demo-10}
\end{figure}



\end{document}